\documentclass{camera}

\usepackage{graphicx}

\begin{document}

\title{A new Multiparametric Topological method for determining the primary
 cosmic ray mass composition in the knee energy region}

\footnote{E-mail address: Michelangelo.Ambrosio@na.infn.it} 

\author{M.Ambrosio~$^{(1)}$, C.Aramo~$^{(1,2)}$, D.D'Urso~$^{(3)}$, A.D.Erlykin~$^{(1,4)}$, F.Guarino~$^{(1,2)}$, A.Insolia~$^{(3)}$}

\organization{$^{(1)}$ INFN, section of Napoli, Napoli, Italy\\
$^{(2)}$ Dept. of Physical Sciences, Univ. of Napoli Federico II, Napoli, Italy\\
$^{(3)}$ Dept. of Physics and Astronomy, Univ. of Catania and INFN, Section of Catania,
 Catania, Italy\\
$^{(4)}$ P.N.Lebedev Physical Institute, Moscow, Russia}

\maketitle

\begin{abstract}
\noindent The determination of the primary cosmic ray mass
composition from the characteristics of extensive air showers (EAS), obtained at 
an observation level in the lower half of the atmosphere, is still an open problem. 
In this work we propose a new method of  
Multiparametric Topological Analysis  and show its applicability for the determination 
of the mass composition of the primary cosmic rays at the PeV 
energy region.
\end{abstract}

\maketitle

\section{Introduction}
It is clear that due to large fluctuations in the longitudinal development
of extensive air showers (EAS) and relatively weak sensitivity of EAS characteristics
to the mass of the parent particle, the observed parameters of showers 
initiated by particles with different primary masses largely overlap. In order to 
estimate the primary mass in the case of the individual shower or to estimate the mean 
mass composition at a certain energy we have to use as many observables as possible. 
To process the big amount of information various multiparametric and non-parametric
methods are employed. Among them are multivariate fitting, KNN-method, Bayesian 
approach, neural net analysis and others. In \cite{Ambr1} we proposed a new
method of the multiparametric topological analysis (MTA) for the study of the 
{\em mean} mass composition on the basis of the EAS longitudinal development data.
We tested this method using the data simulated in the EeV energy region for the Pierre 
Auger experiment and were encouraged by its accuracy and the easy applicability.
In this contribution we test this method at the lower, PeV energy region close to the 
well known 'knee' in the primary cosmic ray energy spectrum ($\sim$3 PeV). Experimental
 arrays for the EAS study in this energy region are usually more compact and are able 
to measure the total size of electron $N_e$, muon $N_\mu$, hadron $N_h$ and Cherenkov 
light $N_{ph}$ components of the shower. The typical example of such a complex EAS 
array is KASCADE \cite{Anton}. Here we use only two parameters: $N_e$ and $N_\mu$, 
although the method is easily generalized for the larger number of observables.
  
\section{The simulated data}

In what follows, we assume that the primary energy
estimate for the observed events is accurate at a few percent level.
In fact, the primary energy estimator for KASCADE is $N_\mu^{tr}$ - the total number of
low energy ($>0.23$ GeV) muons collected in the interval of distances between 40 and 
200 m from the shower core. It has been shown that this number depends just on the 
primary energy and not on the primary mass. The accuracy of the primary energy estimate
in the knee region made with this parameter is determined mainly by the $N_\mu^{tr}$
reconstruction error and is about 5\% \cite{Weber}

The data set used to implement and test the new method consists of 20000 
vertical cascades produced by particles with the fixed energy of 0.5
PeV, 8000 cascades for 1 PeV and 2000 cascades for 5 PeV, simulated using the CORSIKA 
program (version 6.023, \cite{Heck}) with the QGSJET interaction model.
Simulations were performed using Naples PCs .
The primary nuclei were $P$, $He$, $O$ and $Fe$, each of them initiating 5000, 2000 and
 500 cascades for 0.5, 1 and 5 PeV primary energy respectively. The CORSIKA output 
provides various characteristics of the shower: the total number of different particles,
 their lateral distribution and arrival times at different observation levels.

\section{Multiparametric Topological Analysis (MTA) }

\begin{figure}[htb]
\begin{center}
\includegraphics[width=6.2cm]{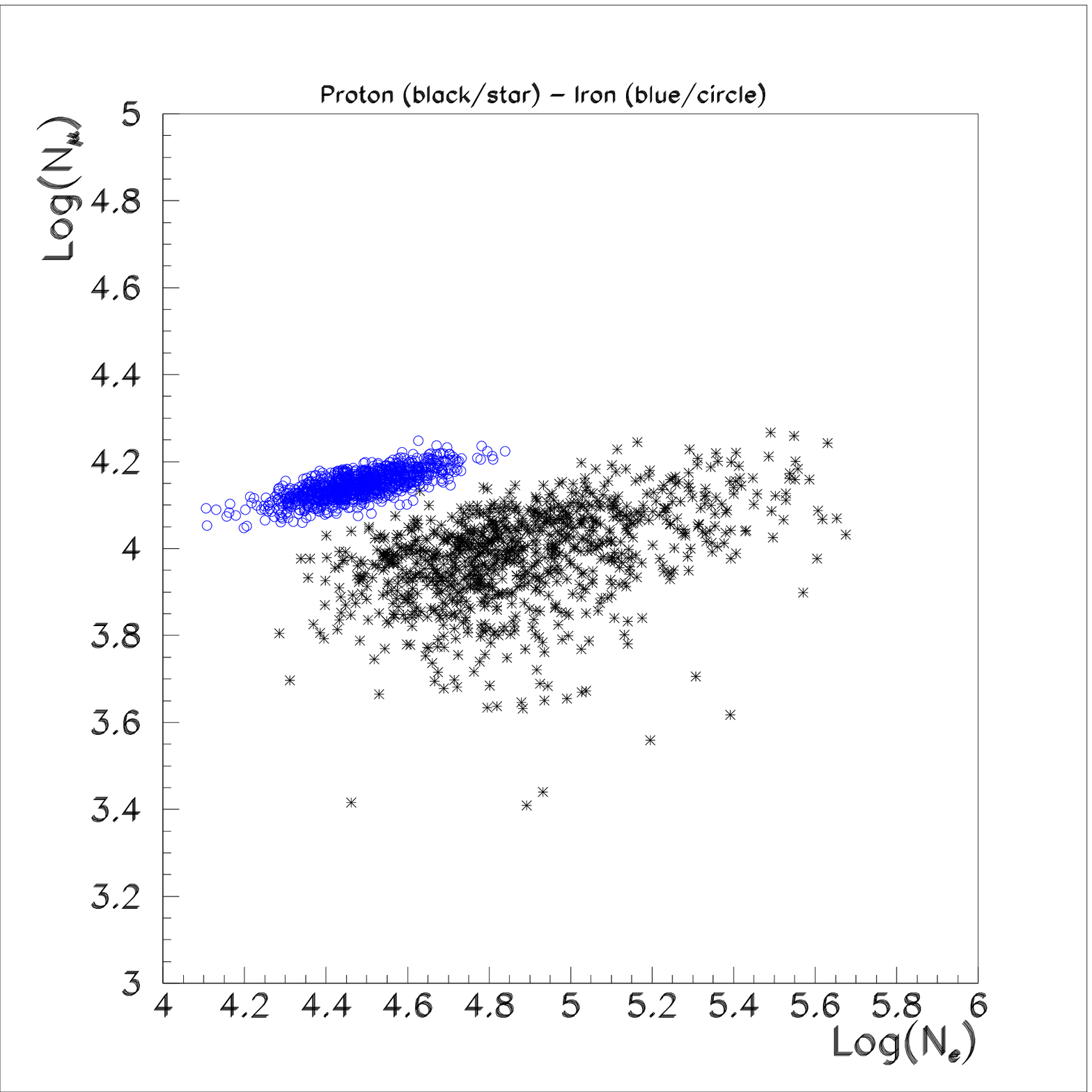}
\includegraphics[width=6.2cm]{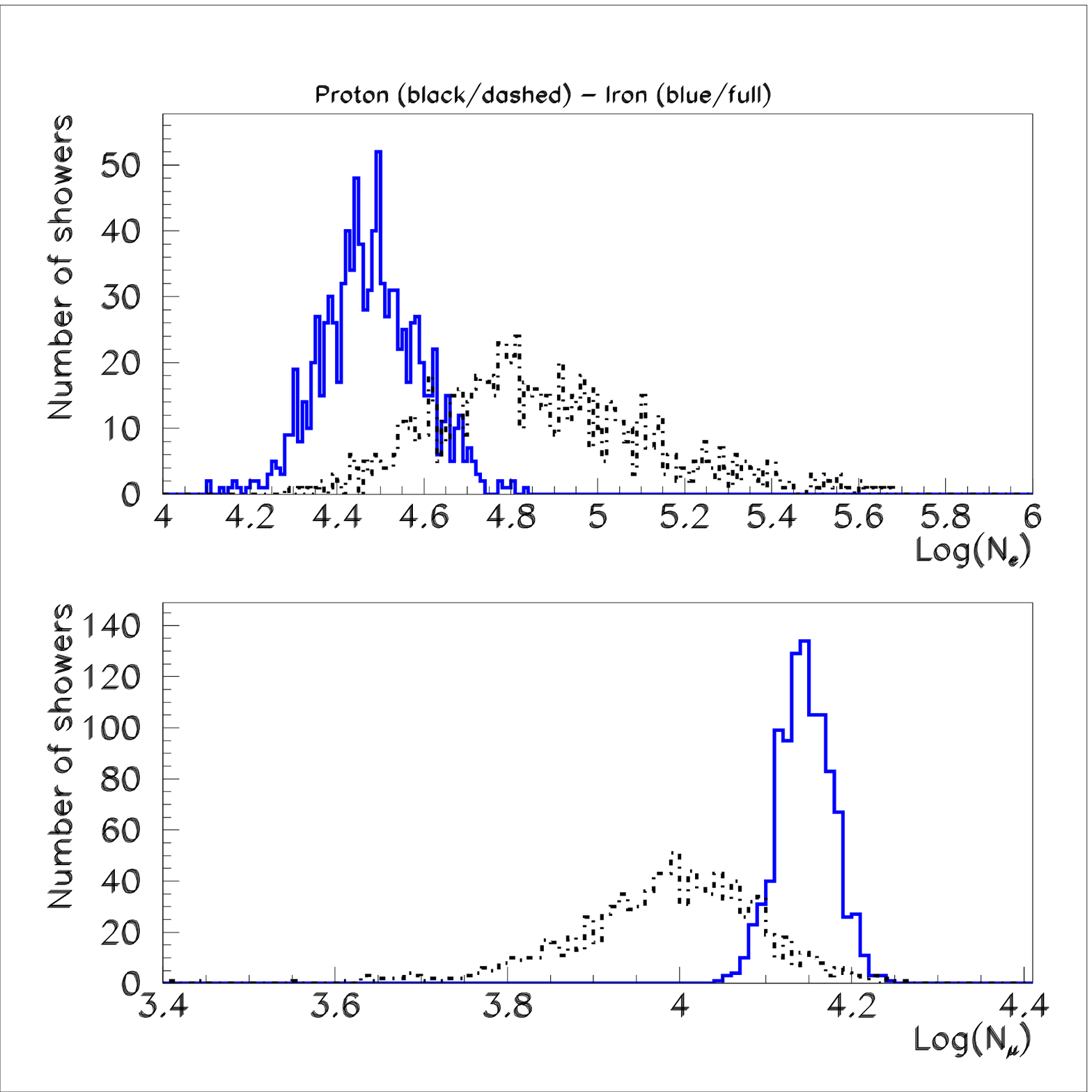}
\caption{\footnotesize (a) $logN_\mu$ vs. $logN_e$ scatter plot for
proton (star) and iron (circle) induced showers at 1 PeV primary energy.
b) Projected distributions of the scatter plot
in (a) for protons (dashed line) and iron nuclei (full line)}
\label{fig:mta1}
\end{center}
\end{figure}

The MTA method, when applied to the simulated data, relies on a
topological analysis of correlations between the most significant
parameters of the shower development in the atmosphere.
In principle this method could be used also with a greater number of parameters,
however, in this paper we restrict ourself to the simple case of two parameters only: 
$logN_e$ and $logN_\mu$ - the total number of electrons and muons at the sea level. 
A scatter plot of these two parameters has been built using the
$4 \times 4000$ showers for 0.5 PeV, $4 \times 1000$ for 1 PeV and 
$4 \times 400$ for 5 PeV primary energy. Figure \ref{fig:mta1}
shows the scatter plot for only proton and iron induced showers of 1 PeV primary energy as well as their projected distributions.

\begin{figure}[t]
\begin{center}
\includegraphics[height=10cm]{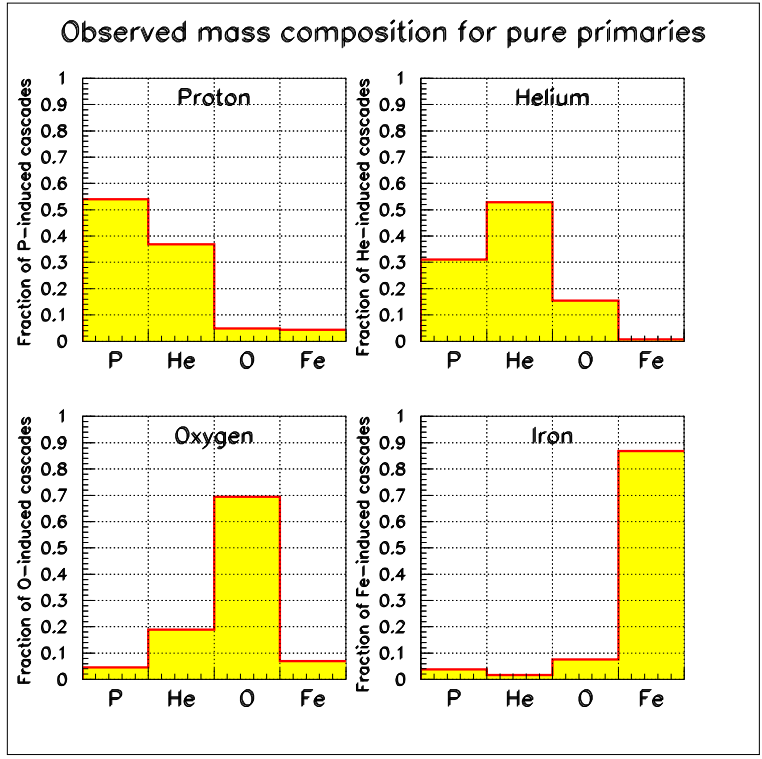}
\caption{\footnotesize Application of the MTA method to the $logN_\mu -
logN_e$ scatter plot - mean probability $P_{ij}$ for the cascades induced by  
{\em i} nucleus: P, He, O and Fe (specified in headers) being identified as 
induced by {\em j} nucleus, indicated at abscissa, at E = 1 PeV.}  
\label{fig:mta3}
\end{center}
\end{figure}

It can be seen that the populations arising from two nuclei are
quite well separated in the $logN_\mu - logN_e$ plane, but not in the projected
one-dimensional plots. The scatter plot of four primary nuclei: $P, He, O$ and $Fe$ 
shows much less separation even in the two-dimensional plot.  The idea of the MTA 
method is 
to divide the entire area of the scatter plot into cells whose dimensions are defined 
by the accuracy with which the parameters can be measured. For this work, we adopted
the accuracy and the size of the cell equal to $\Delta(logN_\mu) = 0.05$ and $
\Delta(logN_e) = 0.02$, which corresponds to the experimental errors of KASCADE array
\cite{Anton}. In each cell we can find the total number
of showers N$_{tot}^{i}$, as the sum of N$_P^{i}$, N$_{He}^{i}$,
N$_O^{i}$ and N$_{Fe}^{i}$ showers induced by P, He, O and Fe
respectively, and then derive the associated frequencies:
p$_P^{i}$=N$_P^{i}$/N$_{tot}^{i}$, p$_{He}^{i}$=N$_{He}^{i}$/N$_{tot}^{i}$, 
p$_O^{i}$=N$_O^{i}$/N$_{tot}^{i}$ and
p$_{Fe}^{i}$=N$_{Fe}^{i}$/N$_{tot}^{i}$ which can be interpreted
as the probability for a real shower falling into the $i^{th}$ cell to
be initiated by proton, helium, oxygen or iron primary nuclei.
In other words, in the case of an experimental data set of $N_{exp}$
showers, it may be seen as composed of a mixture of N$_{exp}\times$
p$_P$ proton showers, N$_{exp}\times$ p$_{He}$ helium showers,
N$_{exp}\times$ p$_O$ oxygen showers and N$_{exp}\times$ p$_{Fe}$ iron
induced showers, where p$_A = \Sigma_i$p$_A^i$. As a probe set 
of 'experimental' showers 
4 subsets of showers, different from those used for the determination of 
probabilities p$_A$, have been used. Each subset consists of 500 showers for 0.5 PeV,
200 showers for 1 PeV and 50 showers for 5 PeV primary energy.
For each individual shower in a
given subset the partial probabilities p$_P^i$ , p$_{He}^i$ , p$_O^i$
and p$_{Fe}^i$ have been read from the relevant cell {\em i}. The sum of
such probabilities over the entire subset permits
to estimate the probability for a shower of a given nature to be
identified as a shower generated by the $P$, $He$, $O$ or $Fe$
primary particle. This probability is shown in Figure \ref{fig:mta3} at 
E = 1 PeV. 
One can see that the method works quite well
being able to attribute the highest probability to the correct nuclei.

\section{\bf Determination of the primary mass composition}

The obtained mean probabilities $P_{ij}$ for
cascades induced by type $i$ nuclei to be identified as those
induced by $j$ nuclei for the pure primary mass composition can be used
for the reconstruction of the mixed primary mass composition as the coefficients in
the system of linear equations:

\begin{eqnarray}
\nonumber n^{\prime}_P & = & n_P \cdot P_{P \rightarrow P} + n_{He} \cdot  P_{He \rightarrow P} + n_O \cdot P_{O \rightarrow P} + 
n_{Fe} \cdot P_{Fe \rightarrow P} \\
\nonumber n^{\prime}_{He} & = & n_P \cdot P_{P \rightarrow He} + n_{He} \cdot P_{He \rightarrow He} + n_O \cdot P_{O \rightarrow He} + 
n_{Fe} \cdot P_{Fe \rightarrow He} \\
\nonumber n^{\prime}_O &= &n_P \cdot P_{P \rightarrow O} + n_{He} \cdot P_{He \rightarrow O} +  n_{He} \cdot P_{O \rightarrow O} + 
n_{He} \cdot P_{Fe \rightarrow O} \\
\nonumber n^{\prime}_{Fe} & =& n_P \cdot P_{P \rightarrow Fe} + n_{He} \cdot P_{He \rightarrow Fe} + n_O \cdot P_{O \rightarrow Fe} +
n_{Fe} \cdot P_{Fe \rightarrow Fe} 
\end{eqnarray}
where $n_P$, $n_{He}$, $n_O$ and $n_{Fe}$ are the true numbers, 
which determine the primary mass composition in the sample of 
$N=n_P+n_{He}+n_O+n_{Fe}$ cascades, which are observed as $n^\prime_P$, 
$n^\prime_{He}$, $n^\prime_O$ and $n^\prime_{Fe}$, due to a misclassification. 

In order to invert the problem and to reconstruct the abundances
$n_{A_i}$ in the primary mass composition from the observed
abundances $n^\prime_{A_i}$ and the known probabilities $P_{ij}$, 
we can apply any method capable to solve the inverse
problem taking into account possible errors of the observed
distribution and the constraint:
\begin{equation}
\Sigma_{i=1}^4 n_{A_i} = \Sigma_{i=1}^4 n_{A_i}^\prime = N
\end{equation}
The observed abundances
were simulated using a set of 4$\times$800 cascades different from those 
used for the determination of $P_{ij}$. 
 At the moment the result of the solution
for the uniform primary mass composition at E = 1 PeV, with all abundances 
$p_A= \frac{n_A}{N}$ = 0.25, reconstructed by MTA method is $p_P = 0.26$, $p_{He} = 0.22$, $p_{O} = 0.27$ 
and  $p_{Fe} = 0.25$. 

As an example of the non-uniform mass composition we have taken the case with
$p_P = p_{Fe}$ = 0.1, $p_{He} = p_{O}$ = 0.4, 
which resembles the mass composition of cosmic rays at the knee energy of 
$\sim$3 PeV \cite{Kampe}. The total amount of cascades used for the simulation of 
this distribution is 2000 (200 $P$ + 800 $He$+ 800 $O$ +200 $Fe$). 
The result of the solution for this case is $p_P = 0.14$, $p_{He} = 0.34$, $p_{O} = 0.42$ 
and  $p_{Fe} = 0.10$ at E = 1 PeV .

It is seen that absolute values of the abundance are reconstructed  
satisfactorily.

For the time being the values of 
the probabilities $P_{ij}$ are assumed to be known precisely. A more accurate 
solution of the system, taking propely into account the errors of the 
probability, is under study.

\section{Application for experimental cascades}

It has to be stressed that all the results
outlined above are biased by the fact that we are dealing with
simulated 'noiseless' data and more realistic testing has to be
performed using data which take into account the varying primary
energy and inclination angle, the instrumental signature (noise)
and the various sources of errors (uneven and incomplete sampling,
etc.). Application of MTA is straightforward - the relevant simulations 
should be made with the largest possible statistics. 

\begin{figure}[hbt]
\begin{center}
\includegraphics[width=9.5cm,height=9.5cm]{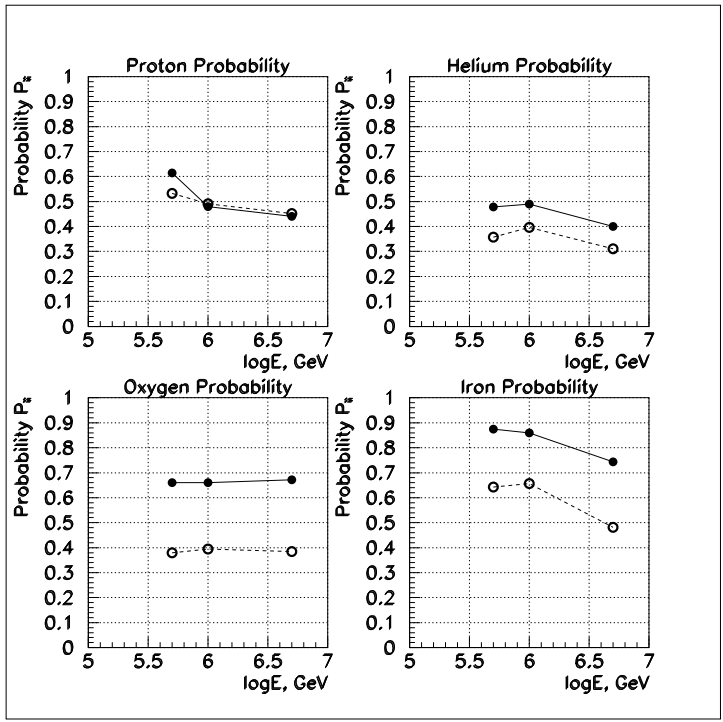}
\caption{\footnotesize The probability of the correct identification of the primary 
nucleus $P_{ii}$ as the function of its energy without (full circles) and with 
(open circles) an account of 
experimental errors: $\Delta(logN_e)$ = 0.02 and $\Delta(logN_\mu)$ = 0.05. The primary
 nuclei are: (a)-P ; (b)-He ; (c)-O ; (d)-Fe.} 
\label{fig:mta6}
\end{center}
\end{figure}

The preliminary study of the effect of the experimental errors shows rather strong
dependence of the probabilities $P_{ij}$ on the adopted errors. In Figure 
\ref{fig:mta6} we show the probabilities $P_{ii}$ for the correct identification of 
nuclei as a function of the primary energy. The experimental accuracies of the shower 
size determination have been taken as $\Delta(logN_e)$ = 0.02 and $\Delta(logN_\mu)$ 
= 0.05. It is seen that while the energy dependence is week the reduction of 
probability $P_{ii}$ is substantial.    

\section{Conclusions}

We proposed and tested the Multiparametric Topological Analysis method for the 
determination of the {\em mean} primary cosmic ray mass composition on the basis of 
measurements of just two parameters of the observed EAS - the total number of electrons 
and muons. Definitely the method needs 
further development before being used for the processing of real experimental cascades,
but the first results are very encouraging. The next efforts should include the 
increase of simulation statistics, the study of the effect of the shower energy 
spectrum and angular distribution, the accurate estimate of the reconstruction errors.
The great advantage of the proposed MTA method is that it is extremely easy to use 
and generalize for the larger number of observables.

\vspace{5mm}

\noindent{\large{\bf Acknowledgements}}
\medskip

\noindent  Authors thank K.H.Kampert and L.Perrone for useful discussions.

\end{document}